\documentclass{article}
\usepackage{amsmath}
\usepackage {amsfonts}
\usepackage {amssymb}

\begin{document}

\begin{center}
\bigskip

\textbf{T.S. GADJIEV, R.A. RASULOV, S.Ya. ALIEV}

\textbf{ON BEHAVIOR OF SOLUTIONS OF DEGENERATED NONLINEAR PARABOLIC EQUATIONS%
}
\end{center}

\bigskip

The aim of this work is studing the behavior of solutions of initial
boundary problem for degenerated nonlinear parabolic equation of the second
order, conditions of existence and non-existence in whole by time solutions,
is establish.

\textbf{1. The exists and nonexists of solutions.} Let's consider the
equation%
\begin{equation*}
\frac{\partial u}{\partial t}=\sum\limits_{i,j=1}^{u}\frac{\partial }{%
\partial x_{j}}\left( \omega \left( x\right) \left\vert \frac{\partial u}{%
\partial x_{i}}\right\vert ^{p-2}\frac{\partial u}{\partial x_{i}}\right)
+f\left( x,t,u\right) .\eqno\left( 1\right)
\end{equation*}%
In bounded domain $\Omega \subset R^{n},$ $n\geq 2$ with nonsmooth boundary,
namely the boundary $\partial \Omega $ contains the conic points with mortar
of the corner $\omega \in \left( 0,\pi \right) $. Denote by $\Pi
_{a,b}=\left\{ \left( x,t\right) :x\in \Omega ,\text{ }a<t<b\right\} ,$ $%
\Gamma _{a,b}=\left\{ \left( x,t\right) :x\in \partial \Omega ,\text{ }%
a<t<b\right\} ,$ $\Pi _{a}=\Pi _{a,\infty }$ $,$ $\Gamma _{a}=\Gamma
_{a,\infty }$. The functions $f\left( x,t,u\right) ,$ $\dfrac{\partial
f\left( x,t,u\right) }{\partial u}$ are continuous by $u$ uniformly in $%
\overline{\Pi }_{0}\times \left\{ u:\left\vert u\right\vert \leq M\right\} $
at any $M<\infty ,$ $f\left( x,t,0\right) \equiv 0,\left. \dfrac{\partial f}{%
\partial u}\right\vert _{u=0}\equiv 0.$ Besides the function $f$ is
measurable on whole arguments and not decrease by $u$. Let's consider the
Dirichlet boundary condition%
\begin{equation*}
u=0,x\in \partial \Omega, \eqno\left( 2\right)
\end{equation*}%
and the initial condition%
\begin{equation*}
\left. u\right\vert _{t=0}=\varphi \left( x\right), \eqno\left(
3\right)
\end{equation*}%
in some domain $\Pi _{0,a}$, where $\varphi \left( x\right) $ is a smooth
function. Further we'll weak this condition.

Solution of problem $\left( 1\right) -\left( 3\right) $ either exist in $\Pi
_{0}$ or%
\begin{equation*}
\lim_{t\rightarrow T-0}\max_{\Omega }\left\vert u\left( x,t\right)
\right\vert =+\infty , \eqno\left( 4\right)
\end{equation*}%
at some $T=const.$

Assuming that $\omega \left( x\right) $ is measurable non-negative function
satisfying the conditions: $\omega \in L_{1,loc}\left( \Omega \right) $ and
for any $r>0$ and some fixed $\theta >1$%
\begin{equation*}
\int\limits_{B_{r}}\omega ^{-1/\left( \theta -1\right) }dx<\infty
,\ \underset{x\in B_{r}}{ess\sup }\omega \leq c_{1}r^{n\left(
\theta -1\right) }\left( \int\limits_{B_{r}}\omega ^{-1/\left(
\theta -1\right) }dx\right) ^{1-\theta },\eqno\left( 5\right)
\end{equation*}
here $B_{r}=\left\{ x\in \Omega :\left\vert x\right\vert <r\right\} .$

From condition $\left( 5\right) $ it follows that%
\begin{equation*}
\underset{x\in \Omega _{r}}{ess\sup }\omega \left( x\right) \leq
c_{1}r^{-n}\int\limits_{B_{r}}\omega dx , \eqno\left( 6\right)
\end{equation*}%
and $\omega \in A_{\theta }$ i.e.%
\begin{equation*}
\int\limits_{B_{r}}\omega dx\left[ \int\limits_{B_{r}}\omega
^{-1/\left( \theta -1\right) }dx\right] ^{1-\theta }\leq
cr^{n\theta } . \eqno\left( 7\right)
\end{equation*}%
Condition $\left( 6\right) -\theta $ is Makenkhoupt's condition (see [3]).

Besides, analogously to [1] we'll assume that $\omega \in D_{\mu },\mu
<1+p/n $, i.e.%
\begin{equation*}
\frac{\omega \left( B_{s}\right) }{\omega \left( B_{h}\right)
}\leq c_{1}\left( \frac{s}{h}\right) ^{n\mu } , \eqno\left(
8\right)
\end{equation*}%
for any $S\geq h>0,$ where $\omega \left( B_{s}\right)
=\int\limits_{B_{s}}\omega \left( x\right) dx.$

Introduce the Sobole's weight space $W_{p}^{1},W_{p,\omega }^{1}\left(
\Omega \right) $ with finite norm%
\begin{equation*}
\left\Vert u\right\Vert _{W_{p,\omega }^{1}\left( \Omega \right)
}=\left( \int\limits_{\Omega }\omega \left( x\right) \left(
\left\vert u\right\vert ^{p}+\left\vert \nabla u\right\vert
^{p}\right) dx\right) ^{1/p}.
\end{equation*}%
The generalized solution of problem $\left( 1\right) -\left( 3\right) $ in $%
\Pi _{0,a^{\prime }}$ we'll call the function $u\left( x,t\right) \in
W_{p,\omega }^{1}\left( \Pi _{a,b}\right) $, such that%
\begin{equation*}
\int\limits_{\Pi _{a,b}}\psi \frac{\partial u}{\partial t}%
dxdt+\sum\limits_{i,j=1}^{n}\int\limits_{\Pi _{a,b}}\omega \left(
x\right) \left\vert \frac{\partial u}{\partial x_{i}}\right\vert
^{p-2}\frac{\partial u}{\partial x_{i}}\frac{\partial \psi
}{\partial x_{j}}dxdt=
\end{equation*}%
\begin{equation*}
=\int\limits_{\Pi _{a,b}}f\left( x,t,u\right) \psi \left( x,t\right) dxdt ,%
\eqno\left( 9\right)
\end{equation*}%
where $\psi \left( x,t\right) $ is an arbitarary function from $W_{p,\omega
}^{1}\left( \Pi _{a,b}\right) ,\left. \psi \right\vert _{\Gamma _{a,b}}=0,$ $%
0<a<b$ are any numbers.

Let's formulate some auxillary result's from [3],[4]. For this
we'll determine $p-$harmonic operator $L_{p}u={div}\left(
\left\vert \nabla u\right\vert ^{p-2}\nabla u\right)$, $p>1$.

\textbf{Lemma 1.} ([1]). \textit{There exists positive eigenvalue of
spectral problem for operator }$L_{p}$\textit{\ that corresponds the
positive in }$\Omega $\textit{\ eigenfunction.}

\textbf{Lemma 2.} ([2]). \textit{Let }$u,v\in W_{p}^{1}\left( \Omega \right)
,u\leq v$\textit{\ on }$\partial \Omega $\textit{\ and}%
\begin{equation*}
\int\limits_{\Omega }L_{p}\left( u\right) \eta _{xi}dx\leq
\int\limits_{\Omega }L_{p}\left( \vartheta \right) \eta _{x_{i}}dx
,
\end{equation*}%
\textit{for any }$\eta \in \overset{\circ }{W}_{p}^{1}\left( \Omega \right) $%
\textit{\ with }$\eta \geq 0$\textit{. Then }$u\leq \vartheta $\textit{\ on
all domain }$\Omega .$

Let $u_{0}\left( x\right) >0$ be an eigenfunction of spectral problem for
the operator $L_{p}$ corresponding $\lambda =\lambda _{1}>0,$ $%
\int\limits_{\Omega }u_{0}\left( x\right) dx=1.$

Let's assume that the condition:%
\begin{equation*}
I=\int\limits_{\Omega }\omega \left( x\right) \left( \left\vert \frac{%
\partial u}{\partial x_{i}}\right\vert ^{p-2}\frac{\partial u}{\partial x_{i}%
}-\left\vert \frac{\partial u_{0}}{\partial x_{i}}\right\vert ^{p-2}\frac{%
\partial u_{0}}{\partial x_{i}}\right) \frac{\partial \left( u_{0}\omega
\right) }{\partial x_{i}}dx\geq 0\eqno\left( \ast \right)
\end{equation*}%
be fulfilled.

\textbf{Theorem 1.} \textit{Let }$f\left( x,t,u\right) \geq \alpha
_{0}\left\vert u\right\vert ^{\sigma -1}u$\textit{\ at }$\left( x,t\right)
\in \Pi _{0},u\geq 0,$\textit{\ where }$\sigma =const>1,\alpha _{0}=const>0$%
\textit{. There exists }$k=const>0$\textit{\ such that if
}$u\left( x,0\right) \geq 0,$\textit{\ }$\int\limits_{\Omega
}u\left( x,0\right)
u_{0}\left( x\right) dx\geq k,$ \textit{and\ condition (*) be fulfilled, then%
}%
\begin{equation*}
\underset{t\rightarrow T-0}{\lim }\underset{\Omega }{\max }\left( \omega
\left( x\right) u_{0}\left( x\right) u\left( x,t\right) \right) =\infty ,
\end{equation*}%
\textit{where }$T=const>0$\textit{.}

\textbf{Proof.} Let's assume the opposite. Then $u\left( x,t\right) $ is a
solution of equation (1) in $\Pi _{0}$ and condition (2) on $\Gamma _{0}$ be
fulfilled. By means of lemma 2 $u\left( x,t\right) >0$\ in $\Pi _{0}$.
Substituts in (8) $\Psi =\varepsilon ^{-1}u_{0}\left( x\right) \omega \left(
x\right) ,$ $b=a+\varepsilon ,$ $a>0,$ $\varepsilon >0,$ where $u_{0}\left(
x\right) >0$ in $\Omega $\ is eigenfunction of spectral problem for the
operator $L_{p}$ corresponding to eigenvalue $\lambda _{1}>0$.\ Such
eigenvalue exists by virtue of lemma 1.

As a result we'll obtain%
\begin{equation*}
\varepsilon ^{-1}\left[ \int\limits_{\Omega }\omega \left(
x\right) u_{0}\left( x\right) u\left( x,a+\varepsilon \right)
dx-\int\limits_{\Omega }\omega \left( x\right) u_{0}\left(
x\right) u\left( x,a\right) dx\right] +
\end{equation*}%
\begin{equation*}
+\varepsilon ^{-1}\int\limits_{\Pi _{a,a+\varepsilon }}\omega
\left(
x\right) \left\vert \frac{\partial u}{\partial x_{i}}\right\vert ^{p-2}\frac{%
\partial u}{\partial x_{i}}\frac{\partial \psi }{\partial x_{j}}%
dxdt=\varepsilon ^{-1}\int\limits_{\Pi _{a,a+\varepsilon
}}u_{0}\omega f\left( x,t,u\right) dxdt . \eqno\left( 10\right)
\end{equation*}

Let's make same transformations. Let's add and substract to left hand (10)%
\begin{equation*}
\varepsilon ^{-1}\int\limits_{\Pi _{a,a+\varepsilon }}\omega
\left(
x\right) \left\vert \dfrac{\partial u_{0}}{\partial x_{i}}\right\vert ^{p-2}%
\dfrac{\partial u_{0}}{\partial x_{i}}\dfrac{\partial \psi }{\partial x_{j}}%
dxdt,
\end{equation*}%
and taking into account that $u_{0}\left( x\right) $ the egenfunction of the
operator $L_{p}$ corresponds to $\lambda _{1}>0$ and $\varepsilon $ vanich
we'll obtain that at all $t>0$%
\begin{equation*}
\frac{\partial }{\partial t}\int\limits_{\Omega }u_{0}\left(
x\right) \omega \left( x\right) u\left( x,t\right) dx=-\lambda
_{1}\int\limits_{\Omega }u_{0}\left( x\right) \omega \left(
x\right) u\left( x,t\right) dx+\int\limits_{\Omega }u_{0}\omega
\left( x\right) f\left( x,t,u\right) dx+I.
\end{equation*}%
From here denoting%
\begin{equation*}
g\left( t\right) =\int\limits_{\Omega }u_{0}\left( x\right) \omega
\left( x\right) u\left( x,t\right) dx ,
\end{equation*}%
we have%
\begin{equation*}
g^{\prime }\left( t\right) =\lambda _{1}\int\limits_{\Omega
}u_{0}\left( x\right) \omega \left( x\right) u\left( x,t\right)
dx+I+\int\limits_{\Omega }u_{0}\omega f\left( x,t,u\right) .
\end{equation*}%
Further, taking into account condition (A) and condition on $f\left(
x,t,u\right) $ we have%
\begin{equation*}
g^{\prime }\left( t\right) \geq -\lambda _{1}\int\limits_{\Omega
}u_{0}\omega \left( x\right) u\left( x,t\right)
dx+a_{0}\int\limits_{\Omega }u_{0}\omega \left\vert u\right\vert
^{\sigma }dx .\eqno\left( 11\right)
\end{equation*}%
So, from (10) we'll obtain%
\begin{equation*}
g^{\prime }\left( t\right) \geq -\lambda _{1}\int\limits_{\Omega
}\omega uu_{0}dx+a_{0}\int\limits_{\Omega }u_{0}\omega u^{\sigma
}dx . \eqno\left( 12\right)
\end{equation*}%
By virtue unequality Holder we have%
\begin{equation*}
\left( \int\limits_{\Omega }uu_{0}\omega dx\right) ^{\sigma }\leq
\left[ \left( \int\limits_{\Omega }u^{\sigma }u_{0}\omega
dx\right) ^{1/\sigma
}\left( \int\limits_{\Omega }\omega u_{0}dx\right) ^{\sigma -1/\sigma }%
\right] ^{\sigma }\leq C_{1}\int\limits_{\Omega }u^{\sigma
}u_{0}\omega dx .
\end{equation*}%
In results%
\begin{equation*}
g^{\prime }\left( t\right) \geq -\lambda _{1}g\left( t\right)
+Cg^{\sigma }\left( t\right) ,\ \ C=const>0 . \eqno\left(
13\right)
\end{equation*}%
If%
\begin{equation*}
g\left( 0\right) >c_{2}=\left( \frac{\lambda _{1}}{c}\right)
^{1/\sigma },
\end{equation*}%
then from (13) we'll obtain $\underset{t\rightarrow T-0}{\lim }g\left(
t\right) =+\infty $. This means that%
\begin{equation*}
\underset{t\rightarrow T-0}{\lim }\underset{\Omega }{\max }\left( \omega
\left( x\right) u_{0}\left( x\right) u\left( x,t\right) \right) =\infty
\end{equation*}

Theorem is proved.

So equation (1) hasn't solutions in satisfying the boundary condition (2) if
$u\left( x,0\right) \geq 0$ isn't much small. Now we'll show that at small $%
\left\vert u\left( x,0\right) \right\vert $ solution of problem (1),(2)
exists on whole domain $\Pi _{0}$.

\textbf{Theorem 2.} \textit{We'll assume that }$\left\vert f\left(
x,t,u\right) \right\vert \leq \left( C_{3}+C_{4}t^{m}\right) \left\vert
u\right\vert ^{\sigma },\ \ \sigma >1,$\textit{\ }$m>1$\textit{. There
exists }$\delta >0$\textit{\ such that if }$\left\vert \varphi \left(
x\right) \right\vert \leq \delta $\textit{\ then solution of problem (1),(3)
exists in }$\Pi _{0}$\textit{\ and }$\left\vert u\left( x,t\right)
\right\vert \leq C_{5}e^{-\alpha ,t},\alpha =const>0$\textit{\ not depend at
}$n$\textit{.}

\textbf{Proof.} Let $\overline{\Omega }\subset B_{R},$ where $%
B_{R}=\{x:\left\vert x\right\vert \leq R\}.$ Let $\vartheta >0$ in $B_{R}$
be eigenfunction corresponding to positive eigenvalue $\lambda _{1}$ of the
boundary problem%
\begin{equation*}
L_{p}u+\lambda u=0,x\in \Omega ,u=0,x\in \partial \Omega .
\eqno\left( 14\right)
\end{equation*}%
Let's consider the function $V\left( x,t\right) =\varepsilon \cdot
e^{-\lambda _{1}t/2}\cdot \vartheta \left( x\right) .$ We have%
\begin{equation*}
\left.
\begin{array}{l}
V_{t}-L_{p}V-f\left( x,t,V\right) =\frac{1}{2}\varepsilon \lambda
_{1}e^{-\lambda _{1}t/2}\cdot \vartheta \left( x\right) - \\
-\left( c_{3}+c_{4}t^{m}\right) \varepsilon ^{\sigma }e^{-\lambda
_{1}t/2}\cdot \vartheta \geq 0,\left( x,t\right) \in \Pi _{0} \\
\text{and }V>0,\left( x,t\right) \in \Gamma _{0},%
\end{array}%
\right. \eqno\left( 15\right)
\end{equation*}%
if $\varepsilon >0$ is sufficiently small. Inequality (15) is understood in
weak sense (see [4]).

From (15) and lemma 2 follows that $\left\vert u\right\vert \leq V\leq
C_{s}e^{-\lambda _{1}t},\ \left\vert \varphi \left( x\right) \right\vert
\leq \delta =\varepsilon \underset{\Omega }{\min }\vartheta \left( x\right) $%
. Let's determine the class of functions $K$ consisting from $g\left(
x,t\right) $ continuous in $\overline{\Pi }_{-\infty ,+\infty }$ equaling to
zero at $t\leq T$ and such that $\left\vert g\left( x,t\right) \right\vert
\leq Ce^{-ht}.$ $K$ is a set of Banach space continuous in $\overline{\Pi }%
_{-\infty ,+\infty }$ functions with norm%
\begin{equation*}
\left\Vert g\right\Vert =\underset{\overline{\Pi }_{-\infty
,+\infty }}{\sup }\left\vert ge^{ht}\right\vert .
\end{equation*}%
Let $\theta \left( t\right) \in C^{\infty }\left( R^{1}\right) ,$ $\theta
\left( t\right) \equiv 0,$ $t\leq T,$ $\theta \left( t\right) =1,$ $t>T+1$.
Let's determine the operator $H$ on $K$ puthing $Hg=\theta \left( t\right) z,
$ $g\in K,$ where $z$ is a solution of lineazing problem.

By virtue of above obtained estimation $H$ transforms $K$ in $K$ if $T$ is
sufficiently big. The operator $H$ is a fully continuous. This follows from
the obtained estimation and theorem on Holderness of solutions of parabolic
equations in $\Pi _{-a,a}$ at any $a$ ([4]). From Lere-Shauder theorem,
consequense that the operator $H$ has fixed point $z$. This shows the
existence of solution.

The theorem is proved.

From theorem 2 it follows that if $u\left( x,0\right) \geq 0,$ $\left\vert
u\left( x,0\right) \right\vert \leq \delta ,$ then the solution of problem
(1)-(3) exists in $\Pi _{0}$ and possitive in $\Pi _{0}$ by virtue of lemma
2.

Let's indicate the sufficient condition, at which all nonnegative solutions
of problem (1)-(3) have "blow-up", i.e.%
\begin{equation*}
\underset{t\rightarrow T-0}{\lim }\underset{\Omega }{\max }\left( \omega
\left( x\right) u_{0}\left( x\right) u\left( x,t\right) \right) =+\infty ,%
\eqno\left( 16\right)
\end{equation*}%
where $T=const>0.$

\textbf{Theorem 3.} \textit{Let }$f\left( x,t,u\right) \geq C_{6}e^{\lambda
_{1}\sigma t}u^{\sigma }$\textit{\ at }$\left( x,t\right) \in \Pi _{0},$%
\textit{\ }$u\geq 0,$\textit{\ }$\sigma =const>1,$\textit{\ }$\lambda _{1}$%
\textit{\ be positive eigenvalue of problem (14) in }$\Omega $\textit{\ that
corresponds to the possitive in }$\Omega $\textit{\ eigenfunction. If }$%
u\left( x,0\right) \geq 0,$\textit{\ }$u\left( x,0\right) \not\equiv 0,$%
\textit{\ where }$u\left( x,t\right) $\textit{\ is solution of problem
(1)-(3), then it holds (16).}

\textbf{Proof.} Similarly how it has been established by inequality (13)
we'll obtain%
\begin{equation*}
g^{\prime }\left( t\right) \geq -\lambda _{1}g+C_{7}e^{\lambda
_{1}\sigma t}g^{\sigma }\left( t\right) , \eqno\left( 17\right)
\end{equation*}%
where%
\begin{equation*}
g\left( t\right) =\int\limits_{\Omega }\omega \left( x\right)
u_{0}\left( x\right) u\left( x,t\right) dx .
\end{equation*}%
Let $g\left( t\right) =\psi \left( t\right) e^{\lambda _{1}t}.$ From (17) if
follows that $\psi ^{\prime }\geq C_{8}\psi ^{\sigma }.$ Hence $\psi \left(
t\right) \rightarrow +\infty $ at $t\rightarrow T-0.$ Thus $g\left( t\right)
$ tends so $+\infty $ at $t\rightarrow T-0.$ Consequently $\underset{\Omega }%
{\max }\left( \omega \left( x\right) u_{0}\left( x\right) u\left( x,t\right)
\right) $ is also tends to infinity

Theorem is proved.

From theorem 3 we can obtain the following property of solutions of equation
(1)

\textbf{Corollary: }Let $f\left( x,t,u\right) \geq C_{8}e^{\lambda
_{1}\sigma t}u^{\sigma }$ and at $\left( x,t\right) \in \Pi _{0},$ $u\geq 0$
where $\sigma >1$. Then there isn't positive in $\Pi _{0}$ solutions of
equation (1).

\textbf{2. The estimation of solutions.} We'll obtain the estimations for
solutions of problem (1)-(3) in case $f\left( x,t,u\right) =0$ in ternus to
characterising on infinity of initial and weight functions, without a
lower's condition on initial function.

Assume, that $\varphi \left( x\right) \in L_{1}\left( \Omega \right) $.
Denote by $k=n\left( p-1-\mu \right) +p,r>0$ fixed number. Let's consider
the following initial characteristics for $u\left( x,t\right) $ and $\varphi
\left( x\right) $%
\begin{equation*}
\varphi _{r}\left( t\right) =\sup_{\tau \in \left( 0,t\right) }\sup_{\rho
\geq r}\left( \frac{\omega \left( B_{\rho }\right) }{\rho ^{n+p}}\right)
^{1/\left( p-2\right) }\cdot \left\Vert u\left( x,\tau \right) \right\Vert
_{L_{\infty }\left( B_{\rho }\right) },
\end{equation*}%
\begin{equation*}
\left\vert \left\Vert u\left( x,\tau \right) \right\Vert
\right\vert _{r}=\sup_{\rho \geq r}\rho ^{-k/\left( p-2\right)
}\left[ \frac{\omega \left( B_{\rho }\right) }{\rho ^{n\cdot \mu
}}\right] ^{1/\left( p-2\right) }\int\limits_{B_{\rho }}u\left(
x,\tau \right) dx,
\end{equation*}%
\begin{equation*}
\left\vert \left\Vert u\left( x,0\right) \right\Vert \right\vert
_{r}=\left\Vert \varphi \right\Vert _{r}.
\end{equation*}

Let's rewrite the definition of generalized solution (9) in the following
form:%
\begin{equation*}
\int\limits_{\Omega }u\left( x,t\right) \psi \left( x,t\right)
dx+\int\limits_{0}^{t}\int\limits_{\Omega }\left( -u\psi
_{t}+\omega \left\vert \frac{\partial u}{\partial
x_{i}}\right\vert ^{p-2}\frac{\partial u}{\partial
x_{i}}\frac{\partial \psi }{\partial x_{j}}dxdt\right) =
\end{equation*}

\begin{equation*}
=\int\limits_{\Omega }\varphi \left( x\right) \psi \left(
x,0\right) dx,\ \ \ \forall \ o<t<T . \eqno\left( 18\right)
\end{equation*}%
\textbf{Lemma 3:} \textit{Assume that }$u\left( x,t\right) \in W_{p,\omega
}^{1}\left( \Pi _{a,b}\right) $\textit{\ is a generalized solution of
problem (1)-(3) is initial function }$\varphi \left( x\right) \in
C_{0}^{\infty }\left( \Omega \right) $\textit{. Then the following
estimation is true}%
\begin{equation*}
\left\vert u\left( x,t\right) \right\vert \leq C_{9}\left[ \beta \left(
t\right) \right] ^{\left( n+p-n\left( \mu -1\right) \right) /\lambda }\left[
\frac{\rho ^{n\mu }}{\omega \left( B_{\rho }\right) }\right] ^{n/\lambda }%
\left[ \int\limits_{t/\varphi }^{t}\int\limits_{B_{2\rho
}}u^{p}dxdt\right] ^{\left( p-n\left( \mu -1\right) \right) } ,
\eqno\left( 19\right)
\end{equation*}%
\textit{for }$\forall \ o<t<T,$\textit{\ where }$\beta \left( t\right)
=t^{-n\left( p-2\right) /k}\cdot \varphi _{r}^{p-2}\left( t\right) +t^{-1},$%
\begin{equation*}
\lambda =n\left( 2p-2-p\mu \right) +p^{2}.
\end{equation*}

\textbf{Proof:} Let $f\left( x,t\right) \in L_{\infty }\left(
0,T:L_{s}\left( B_{\rho }\right) \right) \cap L_{p}\left( 0,T:\overset{\circ
}{W}_{p,\omega }^{1}\left( B_{\rho }\right) \right) ,s,p>1$. Using the weigh
multiplicate inequality from $\left[ 3\right] ,$ we obtain the inequality%
\begin{equation*}
\int\limits_{0}^{T}\int\limits_{B_{\rho }}\left\vert f\left(
x,t\right) \right\vert ^{q}dxdt\leq
\end{equation*}%
\begin{equation*}
\leq C_{10}\frac{\rho ^{n\cdot \mu }}{\omega \left( B_{\rho
}\right) }\left( \underset{0<t<T}{ess\sup }\int\limits_{B_{\rho
}}\left\vert f\right\vert ^{s}dx\right) ^{\left( p-n\left( \mu
-1\right) \right) /n}\int\limits_{0}^{T}\int\limits_{B_{\rho
}}\omega \left\vert \nabla f\right\vert ^{p}dxdt , \eqno\left(
20\right)
\end{equation*}%
$q=p+\dfrac{s}{n}\left( p-n\left( \mu -1\right) \right) .$ Let $\rho >0,T>0$
are fixed. Let's consider the sequence $T_{k}=T/2-T/2^{k+1},$ $\rho
_{k}=\rho +\rho /2^{k+1},$ $\overline{\rho }_{k}=\frac{1}{2}\left( \rho
_{k}+\rho _{k+1}\right) ,$ $k=0,1,...$. Denote by $B_{k}=B_{\rho _{k}}\ ,$ $%
\overline{B}_{k}=B_{\rho _{k}},$ $\Pi _{k}\equiv B_{k}\times \left(
T_{k},T\right) ,$ $\overline{\Pi }_{k}\equiv \overline{B}_{k}\times \left(
T_{k+1},T\right) .$

Let $\xi _{k}\left( x,t\right) $ be cutting function in $\Pi _{k}$
satisfying the conditions $\xi _{k}=1,$ $\left( x,t\right) \in \overline{\Pi
}_{k},$ $\left\vert \nabla \xi _{k}\right\vert \leq 2^{k+2}/\rho ,$ $0\leq
\dfrac{\partial \xi _{\kappa }}{\partial t}\leq 2^{k+2}\cdot T$.

Besides, let $\alpha >0,$ $\alpha _{k}=\alpha -\alpha /2^{k+2},$ $%
k=0,1,2,... $

Let's substitute $\psi \left( x,t\right) =\left( u-\alpha _{k}\right)
_{t}^{p-1}\xi _{k}^{p}$ in integral identuty (18). Doing transformation,
analogously [5] we'll obtain%
\begin{equation*}
\sup_{T_{k+1}\leq t\leq T}\int\limits_{\overline{B}_{k}}\upsilon
_{k}^{s}dx+\iint\limits_{\overline{\Pi }_{k}}\omega \left\vert
\nabla \vartheta _{k}\right\vert ^{p}dxdt\leq C_{11}2^{kp}\beta
\left( t\right) \iint\limits_{\overline{\Pi }_{k}}\vartheta
_{k}^{s}dxdt , \eqno\left( 21\right)
\end{equation*}%
where $\vartheta _{k}=\left( u-\alpha _{k}\right) ^{2\left( p-1\right) /p},$
$s=p^{2}/2\left( p-1\right) .$

Estimating the right part (21) using (20) and doing some calculations we'll
obtain%
\begin{equation*}
-\iint\limits_{\overline{\Pi }_{k}}\vartheta _{k+1}^{q}dxdt\leq
\iint\limits_{\overline{\Pi }_{k}}\left\vert \vartheta _{k+1}\xi
_{k}\right\vert ^{q}dxdt\leq C_{12}\frac{\rho ^{n\cdot \mu
}}{\omega \left( B_{\rho }\right) }\times
\end{equation*}%
\begin{equation*}
\times \left\{ \iint\limits_{\overline{\Pi }_{k}}\omega \left\vert
\nabla
\vartheta _{k}\right\vert ^{p}dxd\tau +\frac{2^{kp}}{\rho ^{p}}%
\iint\limits_{\overline{\Pi }_{k}}\omega \vartheta _{k}^{p}dxd\tau
\right\} \left( \sup_{T_{k+1}\leq t\leq
T}\int\limits_{\overline{B}_{k}}\vartheta _{k}^{s}dx\right)
^{\left( p-n\left( \mu -1\right) \right) /n}\leq
\end{equation*}%
\begin{equation*}
\leq C_{12}\frac{\rho ^{n\cdot \mu }}{\omega \left( B_{\rho }\right) }\left[
\beta \left( t\right) \right] ^{1+\left( p-n\left( \mu -1\right) \right) /n}%
\left[ \iint\limits_{\overline{\Pi }_{k}}\vartheta
_{k+1}^{s}dxd\tau \right] ^{1+\left( p-n\left( \mu -1\right)
\right) /n} . \eqno\left( 22\right)
\end{equation*}%
Further, we'll use the following estimation%
\begin{equation*}
mesA_{k+1}=mes\left\{ \left( x,t\right) \in \Pi _{k+1}/u\left( x,t\right)
>\alpha _{n+1}\right\} \leq k^{-p}2^{-\left( k+1\right) p}\iint\limits_{%
\overline{\Pi }_{k}}\vartheta _{k}^{s}dxd\tau . \eqno\left(
23\right)
\end{equation*}%
From (19) the Holder inequality and using estimation (22) we have%
\begin{equation*}
\iint\limits_{\Pi _{k+1}}\vartheta _{k+1}^{q}dxd\tau \leq \left(
\iint\limits_{\Pi _{k+1}}\vartheta _{k+1}^{q}dxd\tau \right)
^{s/q}\left( mesA_{k+1}\right) ^{1-s/q}\leq
\end{equation*}%
\begin{equation*}
\leq C_{13}\alpha ^{-p\left( 1-s/q\right) }\left[ \frac{\rho ^{n\cdot \mu }}{%
\omega \left( B_{\rho }\right) }\right] ^{s/q}\left( B\left( t\right)
\right) ^{\left( \left( n+p-n\left( \mu -1\right) /n\right) \cdot \left(
s/q\right) \right) }\times
\end{equation*}%
\begin{equation*}
\times \left( \iint\limits_{\Pi _{k}}\vartheta _{s}^{k}dxd\tau
\right) ^{\left( 1+\left( p-n\left( \mu -1\right) /n\right) \cdot
\left( s/q\right) \right) }. \eqno\left( 24\right)
\end{equation*}%
Hence, using [4] denoting%
\begin{equation*}
M=C_{13}\left[ \frac{\rho ^{n\cdot \mu }}{\omega \left( B_{\rho }\right) }%
\right] ^{n/\lambda }\cdot \left( \beta \left( t\right) \right)
^{\left( n+p-n\left( \mu -1\right) \right) /n}\left(
\iint\limits_{\Pi _{k}}u^{p}dxd\tau \right) ^{\left( p-n\left( \mu
-1\right) \right) /\lambda }
\end{equation*}%
we'll obtain that $\underset{\Pi _{a,b}}{\sup }u\left( x,t\right) \leq M.$

Lemma 3 is proved.

Denote $\eta \left( t\right) =\underset{\tau \in \left( 0,t\right) }{\sup }%
\eta _{r}\left( \tau \right) =\underset{\tau \in \left( 0,t\right) }{\sup }%
\left\vert \left\Vert u\left( x,\tau \right) \right\Vert
\right\vert _{r}.$

\textbf{Lemma 4.} Let's assume that $u\left( x,t\right) \in W_{p,\omega
}^{1}\left( \Pi _{a,b}\right) $ be generalized solution of problem of
(1)-(3), the initial function $\varphi \left( x\right) \in C_{0}^{\infty
}\left( \Omega \right) .$ Then the estimations%
\begin{equation*}
\varphi _{r}\left( t\right) \leq C_{14}\int\limits_{0}^{t}\tau
^{-n\left( p-2\right) /k}\varphi _{r}^{p-1}\left( \tau \right)
d\tau +C_{15}\left[ \eta
\left( t\right) \right] ^{\left( p-n\left( \mu -1\right) \right) /k}, \eqno%
\left( 25\right)
\end{equation*}%
\begin{equation*}
\eta \left( t\right) \leq C_{16}\left\vert \left\Vert \varphi
\right\Vert \right\vert _{r}+C_{17}\left( \int\limits_{0}^{t}\tau
^{\left( p-n\left( \mu -1\right) /p\alpha \right) -1}\left(
\varphi _{r}\left( \tau \right) \right) ^{\left( p-2/p\right)
}\eta \left( \tau \right) d\tau +\right.
\end{equation*}%
\begin{equation*}
\left. +\int\limits_{0}^{t}\tau ^{\left( \left( p+1/p\alpha
\right) \left( p-n\left( \mu -1\right) -1\right) \right) }\left(
\varphi _{r}\left( \tau \right) \right) ^{\left( p-2\left(
p+1\right) /k\right) }\eta \left( \tau \right) d\tau \right)
\eqno\left( 26\right)
\end{equation*}%
are true.

\textbf{Proof.} Let's estimate the following integrals%
\begin{equation*}
\left[ \frac{\rho ^{n\cdot \mu }}{\omega \left( B_{\rho }\right) }\right]
\tau ^{n/\alpha }\left[ \frac{\omega \left( B_{\rho }\right) }{\rho ^{n+p}}%
\right] ^{1/\left( p-2\right) }\tau ^{\left( -n\left( p-2\right) /\alpha
\right) \left( n+p-n\left( \mu -1\right) \right) /\lambda }\cdot \varphi
_{r}^{\left( p-2\right) \left( \left( n+p-n\left( \mu -1\right) \right)
/\lambda \right) }\times
\end{equation*}%
\begin{equation*}
\times \left( \int\limits_{t/4}^{t}\int\limits_{B_{2\rho
}}^{t}u^{p}dxd\tau \right) ^{\left( p-n\left( \mu -1\right)
\right) /\lambda }\leq \left[ \varphi _{r}\left( t\right) \right]
^{\left( p-2\right) \left( \left( n+p-n\left( \mu -1\right)
\right) /\lambda \right) }\times
\end{equation*}%
\begin{equation*}
\times \left( \int\limits_{0}^{t}\tau ^{-n\left( p-2\right)
/\alpha }\varphi _{r}^{p}\left( \tau \right) d\tau \right)
^{\left( p-n\left( \mu -1\right) \right) /\lambda }\leq
C_{18}\varphi _{r}\left( t\right) +\left( \eta \left( t\right)
\right) ^{\left( p-n\left( \overline{\omega }\right) \right)
/\alpha },\eqno\left( 27\right)
\end{equation*}%
\begin{equation*}
\left[ \frac{\rho ^{n\cdot \mu }}{\omega \left( B_{\rho }\right) }\right]
^{n/\lambda }\tau ^{n/\alpha }\left[ \frac{\omega \left( B_{\rho }\right) }{%
\rho ^{n+p}}\right] ^{1/\left( p-2\right) }\tau ^{-\left(
n+p-n\left( \mu -1\right) \right) /\lambda }\left(
\int\limits_{t/4}^{t}\int\limits_{B_{2s}}^{t}u^{p}dxd\tau \right)
\leq
\end{equation*}%
\begin{equation*}
\leq C_{19}\left( \varphi _{r}\left( t\right) \right) ^{\left( p-1\right)
\left( p-n\left( \mu -1\right) \right) /\lambda }+\left( \eta \left(
t\right) \right) ^{\left( p-n\left( \mu -1\right) \right) /\lambda }\leq
\end{equation*}%
\begin{equation*}
\leq C_{20}\varphi _{r}\left( t\right) +\left( \eta \left(
t\right) \right) ^{\left( p-n\left( \mu -1\right) \right) /\alpha
} . \eqno\left( 28\right)
\end{equation*}

Now multiplying the both parts (19) on $\left[ \dfrac{\omega \left( B_{\rho
}\right) }{\rho ^{n+p}}\right] ^{1/\left( p-2\right) }\tau ^{n/\alpha },\
\tau \in \left( t/4,t\right) ,\ \forall t>0$ and allowing for estimations
(27), (28) we'll obtain estimation (25).

For getting estimation (26) we'll substitute in integral identity (18) $\psi
\left( x,t\right) =\tau ^{1/p}u^{1-2/p}\xi ^{p}$. We'll obtain%
\begin{equation*}
\int\limits_{0}^{t}\int\limits_{B_{2\rho }}\omega \tau ^{1/p}\cdot
\left\vert \nabla u\right\vert ^{p}u^{-2/p}\xi ^{p}dxd\tau \leq
\end{equation*}%
\begin{equation*}
\leq C_{21}\rho ^{-p}\int\limits_{0}^{t}\int\limits_{B_{2\rho
}}\omega \tau ^{1/p}u^{p-2/p}dxd\tau
+C_{22}\int\limits_{0}^{t}\int\limits_{B_{2\rho }}\tau
^{1/p-1}u^{2\left( p-1\right) /p}dxd\tau . \eqno\left( 29\right)
\end{equation*}%
Let's estimate integral of the right in (29). We have%
\begin{equation*}
\rho ^{p}\int\limits_{0}^{t}\int\limits_{B_{2\rho }}\omega \tau
^{1/p}u^{p-2/p}dxd\tau \leq \omega \left( B_{2\rho }\right) \rho
^{-\left( n+p\right) }\int\limits_{0}^{t}\int\limits_{B_{2\rho
}}\tau ^{1/p}u^{p-2/p}dxd\tau \leq
\end{equation*}%
\begin{equation*}
\leq C_{23}\left( \frac{\omega \left( B_{\rho }\right) }{\rho ^{n}}\right)
^{-1/p}\left( \frac{\omega \left( B_{\rho }\right) }{\rho ^{n\cdot \mu }}%
\right) ^{-1/\left( p-2\right) }\rho ^{1+\alpha /\left( p-2\right) }\times
\end{equation*}%
\begin{equation*}
\times \int\limits_{0}^{t}\tau ^{\left( \left( p+1\right) /p\alpha
\right) \left( p-n\left( \mu -1\right) \right) -1}\left( \varphi
_{r}\left( t\right) \right) ^{\left( p-2\right) \left( p+1\right)
/p}\eta \left( \tau \right) d\tau . \eqno\left( 30\right)
\end{equation*}%
The second integral on the right in (29) we'll estimate by the following way%
\begin{equation*}
\int\limits_{0}^{t}\int\limits_{B_{2\rho }}\tau
^{\frac{1}{p}-1}u^{2\left(
p-1\right) /p}dxd\tau \leq \left( \frac{\omega \left( B_{\rho }\right) }{%
\rho ^{n}}\right) ^{-1/p}\left( \frac{\omega \left( B_{\rho }\right) }{\rho
^{n\cdot \mu }}\right) ^{-1/\left( p-2\right) }\rho ^{1+\alpha /\left(
p-2\right) }\times
\end{equation*}%
\begin{equation*}
\times \int\limits_{0}^{t}\tau ^{\left( p-n\left( \mu -1\right)
\right) /p\alpha -1}\left( \varphi _{r}\left( \tau \right) \right)
^{\left( p-2\right) /p}\eta \left( \tau \right) d\tau .
\eqno\left( 31\right)
\end{equation*}

Now, let's substitute in integral identity (18) $\psi \left( x,t\right) =\xi
^{p}\left( x\right) .$ Then we'll obtain%
\begin{equation*}
\int\limits_{B_{2\rho }}u\left( x,t\right) dx\leq
\int\limits_{B_{2\rho }}\varphi \left( x\right) dx+C_{24}\rho
^{-1}\int\limits_{0}^{t}\int\limits_{B_{2\rho }}\omega \left\vert
\nabla u\right\vert ^{p-1}\xi ^{p-1}dxd\tau . \eqno\left(
32\right)
\end{equation*}%
Let's estimate the secong integral on the right in (32). We have%
\begin{equation*}
\int\limits_{0}^{t}\int\limits_{B_{\rho }}\omega \left\vert \nabla
u\right\vert ^{\left( p-1\right) }\xi ^{p-1}dxd\tau \leq \left(
\int\limits_{0}^{1}\int\limits_{B_{2\rho }}\omega \tau ^{1/p}\cdot
\left\vert \nabla u\right\vert ^{p}u^{-2/p}\xi ^{p}dxd\tau \right)
^{\left( p-1\right) /p}\times
\end{equation*}

\begin{equation*}
\times \left( \int\limits_{0}^{t}\int\limits_{B_{2\rho }}\omega
\tau
^{-\left( p-1\right) /p}u^{2\left( p-1\right) /p}dxd\tau \right) ^{1/p} . \eqno%
\left( 33\right)
\end{equation*}%
Taking into account the second multiplies in (33)%
\begin{equation*}
\int\limits_{0}^{t}\int\limits_{B_{2\rho }}\omega \tau ^{-\left(
p-1\right) /p}u^{2\left( p-1\right) /p}dxd\tau \leq
C_{25}\frac{\omega
\left( B_{\rho }\right) }{\rho ^{n}}\int\limits_{0}^{t}\int\limits_{B_{2%
\rho }}\tau ^{1/p-1}u^{2\left( p-1\right) /p}dxd\tau . \eqno\left(
34\right)
\end{equation*}%
Now allowing for estimations (30), (31), (32) in (33) we'll obtain%
\begin{equation*}
\int\limits_{0}^{t}\int\limits_{B_{2\rho }}\omega \left\vert
\nabla u\right\vert ^{p-1}\xi ^{p-1}dxd\tau \leq C_{25}\left(
\frac{\omega \left( B_{\rho }\right) }{\rho ^{n\cdot \mu }}\right)
^{-1/\left( p-2\right) }\rho ^{1+\alpha /\left( p-2\right) }\times
\end{equation*}%
\begin{equation*}
\times \left( \int\limits_{0}^{t}\tau ^{\left( \left( p+1\right)
/p\alpha \right) \left( p-n\left( \mu -1\right) \right) -1}\left(
\varphi _{r}\left( \tau \right) \right) ^{\left( p-2\right) \left(
p+1\right) /p}\eta \left( \tau \right) d\tau +\right.
\end{equation*}%
\begin{equation*}
\left. +\int\limits_{0}^{t}\tau ^{\left( p-n\left( \mu -1\right)
\right) /p\alpha -1}\varphi _{r}^{\left( p-2\right) /2}\left( \tau
\right) \eta \left( \tau \right) d\tau ^{\left( p-1\right)
/p}\right) \times
\end{equation*}%
\begin{equation*}
\times \int\limits_{0}^{t}\tau ^{\left( p-n\left( \mu -1\right)
\right) /p\alpha -1}\left( \varphi _{r}\left( \tau \right)
^{\left( p-1\right) /p}\eta \left( \tau \right) d\tau \right)
^{1/p} . \eqno\left( 35\right)
\end{equation*}

Multiplying inequality (32) $\rho ^{-\alpha /\left( p-2\right) }\rho
^{-n\cdot \mu /\left( p-2\right) }\left( \omega \left( B_{\rho }\right)
\right) ^{1/\left( p-2\right) },$ using inequality (35), then we'll obtain%
\begin{equation*}
\eta \left( t\right) \leq C_{27}\left\vert \left\Vert \varphi
\right\Vert \right\vert _{r}+C_{28}\left( \int\limits_{0}^{t}\tau
^{\left( \left( p+1\right) /p\cdot \alpha \right) \left( p-n\left(
\mu -1\right) \right) }\left( \varphi _{r}\left( \tau \right)
^{\left( p-2\right) \left( p+1\right) /p}\eta \left( \tau \right)
d\tau \right) \right) +
\end{equation*}%
\begin{equation*}
+\int\limits_{0}^{t}\tau ^{\left( p-n\left( \mu -1\right) \right)
/p\alpha -1}\left( \varphi _{r}\left( \tau \right) ^{\left(
p-2\right) /2}\eta \left( \tau \right) d\tau \right).
\end{equation*}

Lemma 4 is proved.

\textbf{Theorem 4. }Let $u\left( x,t\right) \in W_{p,\omega }^{1}\left( \Pi
_{a,b}\right) $ be generalized solution of problem (1)-(3) and $\left\vert
\left\Vert \varphi \right\Vert \right\vert _{r}<\infty ,r>0$ be fixed. Then
if relative $\omega \left( x\right) $ to conditions (4), (7) and $\mu <1+p/n$
fulfiled, then%
\begin{equation*}
\left\vert \left\Vert \varphi \right\Vert \right\vert
_{r}<C_{29}t^{1/\left( p-2\right) } , \eqno\left( 36\right)
\end{equation*}%
\begin{equation*}
\left\vert \left\Vert u\left( x,t\right) \right\Vert \right\vert
_{r}<C_{30}t^{1/\left( p-2\right) } , \eqno\left( 37\right)
\end{equation*}%
\begin{equation*}
\sup_{B_{\rho }}\left\vert u\left( x,t\right) \right\vert \leq
C_{31}t^{p\left( n+1\right) -n\left( \mu +1\right) /k\left(
p-2\right) }\rho ^{n+p}\cdot \omega ^{-1}\left( B_{\rho }\right) .
\eqno\left( 38\right)
\end{equation*}

\textbf{Proof:} The proof of theorem follows from lemma 4 usinf the method
of paper [5]. Thus for obtaining estimations (37), (38) the estimations are
at first obtained%
\begin{equation*}
\left\vert \left\Vert u\left( x,t\right) \right\Vert \right\vert
_{r}<C_{32}\left\vert \left\Vert \varphi \right\Vert \right\vert _{r}\ ,
\end{equation*}%
\begin{equation*}
\sup_{B_{s}}\left\vert u\left( x,t\right) \right\vert \leq
C_{33}\left\vert \left\Vert \varphi \right\Vert \right\vert
_{r}^{\left( \rho -n\left( \mu -1\right) \right) /k}\rho
^{n+p}\cdot \omega ^{-1}\left( B_{\rho }\right) t^{-n/k} .
\eqno\left( 39\right)
\end{equation*}
Further, using these estimations we obtain estimations (37), (38)

\textbf{Corollary:} Let in theorem 4 $\omega \left( x\right) =\left\vert
x\right\vert ^{\theta },\ 0<\theta <p$. Then conditions (4), (7) $\mu
=1+\theta /n,$ are fulfilled and we have the following estimation%
\begin{equation*}
\sup_{B_{\rho }}\left\vert u\left( x,t\right) \right\vert \leq
C_{34}\left( \sup_{\rho \geq r}\rho ^{-\beta /\left( p-2\right)
}\int\limits_{B_{\rho }}\varphi \left( x\right) dx\right) ^{\left(
p-\theta \right) /\beta }\cdot
\rho ^{\left( p-\theta \right) /\left( p-2\right) }\cdot t^{-n/\beta } , \eqno%
\left( 40\right)
\end{equation*}%
where $\beta =n\left( p-2\right) +p-\theta$.

Note that estimation (39) is a exactly that proves to be true following
class of exact solutions%
\begin{equation*}
u_{\theta }\left( x,t\right) =\left( 1-\left( \frac{p-2}{p-\theta }\right)
\left( \frac{n}{\beta }\right) ^{1/\left( p-1\right) }\left( \frac{%
\left\vert x\right\vert }{t^{1/\beta }}\right) ^{\left( p-\theta
\right) /\left( p-1\right) }\right) ^{\left( p-1\right) /\left(
p-2\right) }. 
\end{equation*}

In case $\alpha =0$ and considering Cauchy problem estimation (40) is
coinsider with the result of paper [5].

\textbf{Remark:} Estimations of type (38) we can a;so obtain for $\underset{%
B_{\rho }}{\sup }\left\vert \nabla u\left( x,t\right) \right\vert $

\bigskip
\bigskip

\textbf{T.S.Gadjiev}

\bigskip

Institute of Mathematics and Mechanics of NAS of Azerbaijan.

9, F. Agayev str., AZ1141, Baku, Azerbaijan.

Tel.: (99412)

\bigskip
\bigskip

\textbf{S.Ya.Aliev}

\bigskip

Baku State University,Z.Khalilov str.23, Tel.: (99412)5370826 .

9, F. Agayev str., AZ1141, Baku, Azerbaijan.

 \ \ \ \ \ \ \ \ \ \ \ \ \ \ \ \ \ \ \ \ \ \ \ \
\ \ \ \ \ \ \ \ \ \ \ \ \ \ \ \

\end{document}